\documentclass[%
 reprint,
 amsmath,amssymb,
 aps,prapplied
]{revtex4-2}

\usepackage{graphicx}
\usepackage{dcolumn}
\usepackage{bm}
\usepackage{placeins} 

\begin{document}

\title{Galilean Invariance in the Characterization of Light Drag in Moving Atomic Vapor}

\author{Edgar S. Arroyo-Rivera\textsuperscript{1,*}}
\author{Long D. Nguyen\textsuperscript{1}}
\author{Surendar Vijayakumar\textsuperscript{2}}
\author{Akbar Safari\textsuperscript{3,4}}
\author{Robert W. Boyd\textsuperscript{1,2,4}}

\affiliation{\textsuperscript{1}Department of Physics and Astronomy, University of Rochester, Rochester, New York 14627, USA}
\affiliation{\textsuperscript{2}Institute of Optics, University of Rochester, Rochester, New York 14627, USA}
\affiliation{\textsuperscript{3}Department of Physics, University of Wisconsin-Madison, Madison, WI 53706, USA}
\affiliation{\textsuperscript{4}Department of Physics, University of Ottawa, Ottawa, Ontario K1N 6N5, Canada}

\thanks{Corresponding author: earroyor@ur.rochester.edu}

\date{\today}

\begin{abstract}
Light experiences drag effects when it propagates through a moving medium. The study of light drag has provided foundational insights into light-matter interactions. While longitudinal drag has been extensively characterized, transverse drag, where the medium moves perpendicular to the light's propagation, is subtler and requires advanced techniques for detection. In this work, we experimentally investigate transverse drag in a highly dispersive slow-light medium using non-degenerate Zeeman electromagnetically induced transparency (EIT) in rubidium vapor. By systematically comparing configurations where the light beam and the medium serve as the moving frame, we leverage Galilean invariance to analyze transverse light-drag in this optical context. Thus, we provide a platform for future tests of fundamental principles on strong experimental grounds, which offers promising applications in precision velocimetry, accelerometry, quantum information, and light storage technologies.
\end{abstract}

\maketitle

\section{Introduction}

The propagation of light through moving media has been the subject of study for centuries, providing insights into both fundamental physics and applied optics. The concept of light drag, which refers to a change in the velocity of light as it travels through a moving material, was first introduced by Fresnel in 1818~\cite{fresnel_annales_1818} and experimentally verified by Fizeau in 1851~\cite{fizeau_sur_1851,fizeau_xxxii_1860}. In 1877, Lord Rayleigh clearly explained the concept of group velocity ~\cite{ch_dalmeida_rayleigh_1877} and decades later Hendrik Lorentz took a key step forward by implicitly incorporating dispersion effects into the analysis of light-matter interactions~\cite{h_a_lorentz_electromagnetic_1937}, paving the way for modern investigations in moving dispersive media. The light-drag effect has been broadly studied in different configurations: longitudinally, rotary, and transversely ~\cite{akbar_safari_light-drag_2016,ryan_hogan_beam_2023,yakov_solomons_transverse_2020,chitram_banerjee_anomalous_2022,arash_ahmadi_concept_2024}.

We are ultimately interested in studying transverse drag, not only because its implications for optical technologies in developing new approaches to constructing velocimeters and accelerometers, but also because of its relevance in fundamental aspects of light-matter interactions~\cite{mikhail_d_lukin_controlling_2001,yakov_solomons_transverse_2020,chitram_banerjee_anomalous_2022,ryan_hogan_beam_2023,arash_ahmadi_concept_2024}. This phenomenon, manifesting as a lateral displacement of a light-beam, is typically small but can be significantly enhanced in highly dispersive media such as rubidium vapor under conditions of slow light~\cite{akbar_safari_light-drag_2016,phillips_storage_2001}. Here we will use a quantum technique, electromagnetically induced transparency (EIT), to make the transverse drag not only accessible as a tool but also to demonstrate its potential for investigating frame-dependent optical effects in moving media.

While Galilean and Lorentz invariance have been extensively tested in various regimes~\cite{david_mattingly_modern_2005,j_d_tasson_what_2014}, their experimental investigation in the optical domain, particularly in the context of light drag, remain underexplored. Nonlinear optical techniques combined with vapor cell-based EIT experiments work sufficiently well, compared to typical cold atom setups, to access ultraslow group velocities and precise optical control of light beams. For instance, group indices values reaching $n_g = 10^5$ or even $10^7$ could lead to effective velocities exceeding $3.1 \times 10^8$ m/s when the medium moves at $v = 1$ m/s, showcasing the remarkable sensitivity of such systems to relativistic effects. These extreme enhancements make the transverse drag under EIT conditions a unique experimental framework for investigating how light propagation transforms between inertial frames in the presence of slow-light media, particularly under scenarios involving both constant velocities and accelerated motion, as suggested in~\cite{justin_dressel_gravitational_2009}. Our approach builds upon prior studies that demonstrated anomalous drag effects and enhanced light-matter interactions in slow-light systems ~\cite{jonathan_leach_aether_2008,dmitry_strekalov_observation_2004,stanko_n_nikolic_connection_2012,azmat_iqbal_photon_2017} as well as the simplified theoretical model outlined by Solomons et al~\cite{yakov_solomons_transverse_2020}.

In this work, we investigate transverse light drag in rubidium vapor under EIT conditions, focusing on two complementary scenarios: (1) a stationary probe beam with a moving rubidium cell and (2) a stationary rubidium cell with a transversely moving probe beam. By employing Galilean transformations, we demonstrate that the lateral displacement, $\Delta x$, remains consistent between these two scenarios. This investigation provides insight into the transformation properties of transverse light drag in dispersive media and highlights the robustness of this effect under different inertial frames.

\section{Transverse Light-Drag and Slow-light process}
Transverse light-drag effect can be significantly enhanced under slow-light conditions, where the group index $n_g$ is large and inversely related to the group velocity by $n_g = c / v_g$. In our experiment, the moving medium is a 7.5 cm-long and 2 cm-wide glass cell filled with rubidium vapor $(^{87}\text{Rb})$ and 30 Torr of He buffer gas.

\subsection{Light-Drag}
Longitudinal light-drag leads to a change in the phase velocity of light; this change depends on the velocity with which the material moves. In contrast, the transverse drag case, which is the topic of this investigation, leads to a physical displacement at the output of the moving medium as pictured in Fig.~\ref{fig:fig1}(a). For a non-relativistic and constant velocity $v$, the displacement of the outgoing beam can be expressed as~\cite{ma_player_dispersion_1975, iacopo_carusotto_transverse_2003}:

\begin{eqnarray}
\Delta x = L\frac{v}{c} \left(\frac{c}{v_g}-\frac{v_p}{c}\right).
\end{eqnarray}
Here, $L$ is the length of the Rb cell, $c$ is the speed of light in vacuum, $v$ is the velocity of the cell, $v_p$ is the phase velocity, and $v_g$ is the group velocity of light within the cell. When $v_g $ is much smaller than $ c$, the displacement can be approximated as
\begin{equation}
\Delta x \approx \frac{L \nu}{v_g} = \tau \nu.
\end{equation}
The parameter $\tau$ represents the group delay, which is the time it takes for the pulse envelope to traverse the medium. This delay becomes significant in highly dispersive systems such as the three-level atomic gases used in this study.

\subsection{Electromagnetically Induced Transparency}

\begin{figure}
\centering
\includegraphics[width=1.0\linewidth]{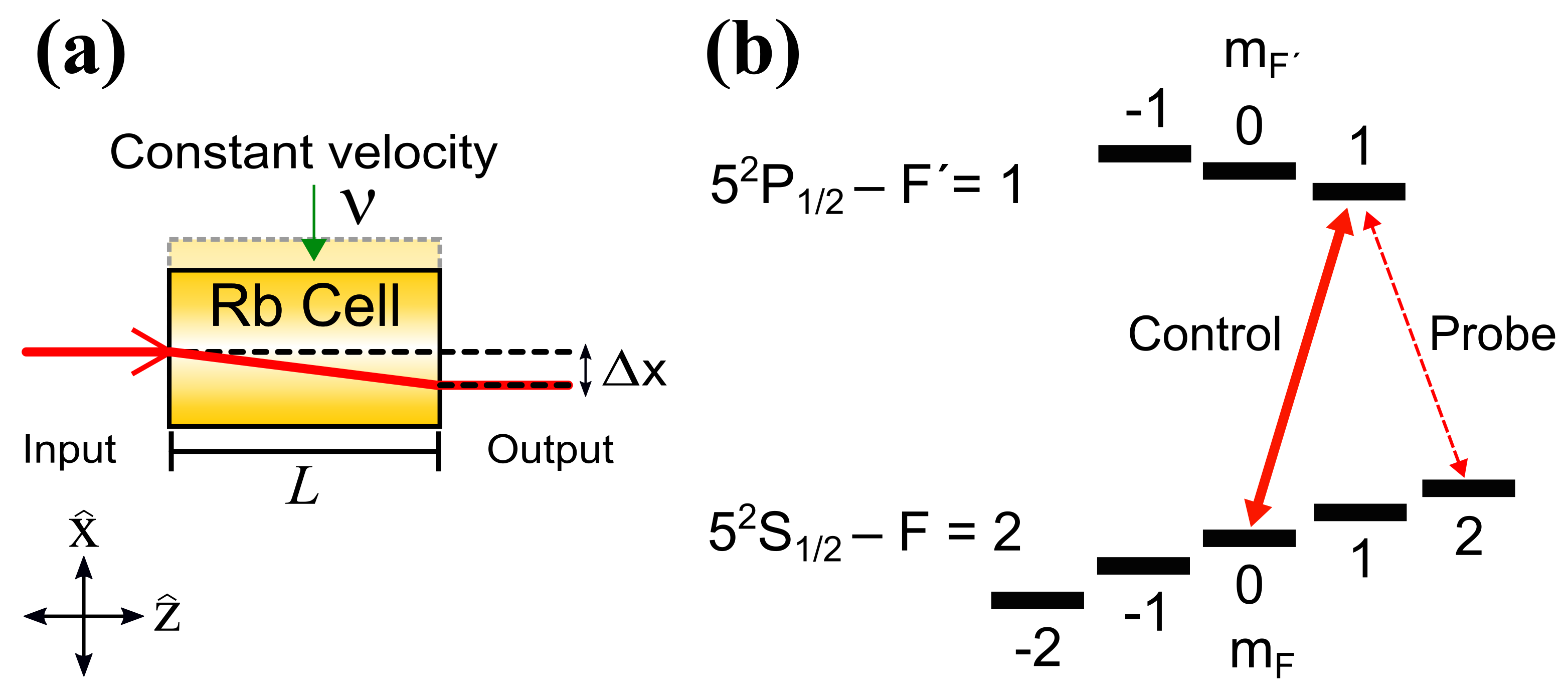} 
\caption{\label{fig:fig1} (a) Schematic representation of the transverse light-drag experiment. The beam experiences a lateral displacement as it propagates through the moving rubidium cell. (b) Non-degenerate Zeeman EIT atomic configuration}
\end{figure}

Electromagnetically induced transparency (EIT)~\cite{atac_imamoglu_observation_1991,michael_fleischhauer_electromagnetically_2005,ran_finkelstein_practical_2023} occurs in a three-level atomic system when a strong control beam and a weak probe beam couple two states of an atomic ground level to a common excited level, as shown in Fig. ~\ref{fig:fig1}(b). Such transition pathways destructively interfere and under specific conditions, this process creates a transparency window for the probe beam that would otherwise be absorbed. However, what is particularly remarkable about EIT, is that near the atomic two-photon resonance, it induces a large and steep dispersion effect at minimal absorption. The two photon resonance arises from the coherent interaction between the probe and control beams, and the steep dispersion effect is mainly responsible of the drastic reduction of the probe beam's group velocity, enabling slow-light effects. 

In our experiment, we implement EIT in a $\Lambda$-type system using the D1 line of $^{87}$Rb. To optimize the slow-light conditions and enhance the EIT resonance, we implemented non-degenerate Zeeman EIT scheme shown in Fig.~\ref{fig:fig1}b. This approach allows precise targeting of specific transitions while minimizing unwanted couplings that can reduce the transparency window.

In the non-degenerate scheme, a magnetic field is applied along the propagation axis, lifting the degeneracy of the Zeeman sub-states. This controlled splitting improves the selectivity of the EIT process by isolating the desired three-level interaction while suppressing undesired transitions. In our experiment, applying a magnetic field of 50 mG induces a Zeeman splitting of approximately 50 kHz, following the Larmor precession relation $\Delta E = g_F \mu_B B$, where $g_F$ is the Landé g-factor, $\mu_B$ is the Bohr magneton, and $B$ is the applied magnetic field. We use the Landé $g$-factor for the $^{87}\mathrm{Rb}$ ground-state hyperfine levels as given in Steck~\cite{d_steck_rubidium_2003}, where $g_F = +\tfrac{1}{2}$ for $F=2$ and $g_F = -\tfrac{1}{2}$ for $F=1$.

\begin{figure*}[t]
\centering
\includegraphics[width=1.0\linewidth]{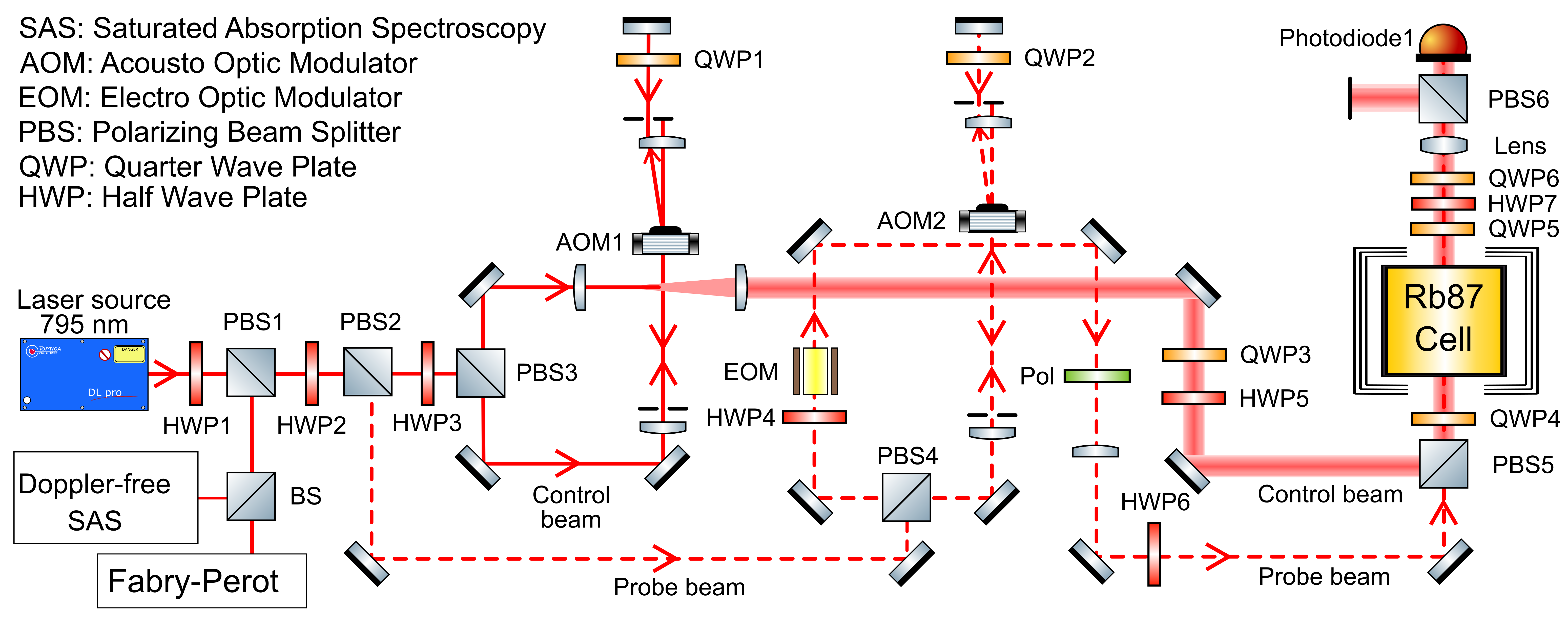}  
\caption{\label{fig:fig2} Experimental setup to generate and measure slow light from non-degenerate Zeeman EIT.}
\end{figure*}

By tuning the frequencies of the control and probe beams accordingly, we ensure efficient EIT conditions with enhanced transparency contrast. After establishing the non-degenerate Zeeman EIT conditions (in which the control and probe fields couple different Zeeman sublevels and therefore have different optical frequencies), we characterized the slow-light properties by scanning the probe frequency and measuring the transmission peak at resonance. The experimental setup used to implement non-degenerate Zeeman EIT is shown in Fig.~\ref{fig:fig2}. The setup is divided into three sections: (1) laser source preparation, (2) control field and probe field configuration, and (3) slow-light medium configuration. 

\textit{Laser source preparation}: The laser source is an external cavity diode laser (Toptica DL Pro), tuned to 795 nm (D1 transition of $^{87}\mathrm{Rb}$~\cite{d_steck_rubidium_2003}). A small portion of the beam is reflected from a beam sampler for Doppler-free saturation absorption spectroscopy (SAS), which allows fine-tuning and frequency locking to the desired transition $\left(5^2 S_{1/2}- F = 2 \leftrightarrow 5^2 P_{1/2}- F' = 1\right)$ ~\cite{d_steck_rubidium_2003}. A scanning Fabry-Perot interferometer is used to verify single-mode operation.

\textit{Control and probe field preparation}: The beam is split at PBS2; the transmitted beam is the control beam and is reflected on PBS3 after polarization rotation via a half-wave plate. Subsequently, it is frequency shifted (80 MHz) and amplitude-modulated by an acousto-optic modulator (AOM1) in a double-pass configuration. This beam is then expanded to a 3-mm diameter to ensure a near-uniform excitation across the 1-mm probe beam. The probe beam also passes through an acousto-optic modulator (AOM2), where it undergoes a frequency shift of 80 MHz. The frequency shift will be swept from 79.5 to 80.5 MHz later to measure the linewidth of the EIT transmission window. An electro-optic modulator (EOM) is used to modulate its amplitude and generate Gaussian pulses for measuring the group delay.

The control and probe beams are recombined on PBS5. QWP3 and HWP5 are used to precisely control the polarization of the control beam so that after the reflection on PBS5, the polarization of the control beam is perfectly orthogonal to that of the probe beam. A 50:50 beam splitter is placed after PBS5 to pick off $50\%$ of each beam for reference purpose. Before entering the Rb cell, the two beams’ polarizations are converted from linear to left and right circular. At the output of the cell, a waveplate array (QWP5-HWP6-QWP6) is used to precisely convert the two beams’ polarizations back to linear and orthogonal. The strong control beam is then rejected by a polarizing beam splitter (PBS) so that only the probe beam falls upon the photodiode. Precise polarization control of the probe and control beams using quarter-wave plates (QWPs) and half-wave plates (HWPs) is essential for ensuring proper coupling according to the dipole selection rules and maximizing the EIT contrast.

\textit{Slow-light-medium configuration}: The Rb vapor cell is magnetically shielded by 4 layers of mu-metal to minimize stray magnetic fields that could perturb the Zeeman sublevels. Before collecting any data, we perform a degaussing procedure by gradually reducing the current through the degaussing coils, located in the two outermost layers of the magnetic shield, from 5 A to 200 mA over 1–2 minutes using two variac devices in series, and then decreasing it further in finer steps down to zero. The magnetic field along the propagation axis is generated by a 4-layer solenoid with about 400 turns of enameled copper wire around the cell holder.

\begin{figure}
\centering
\includegraphics[width=0.8\linewidth]{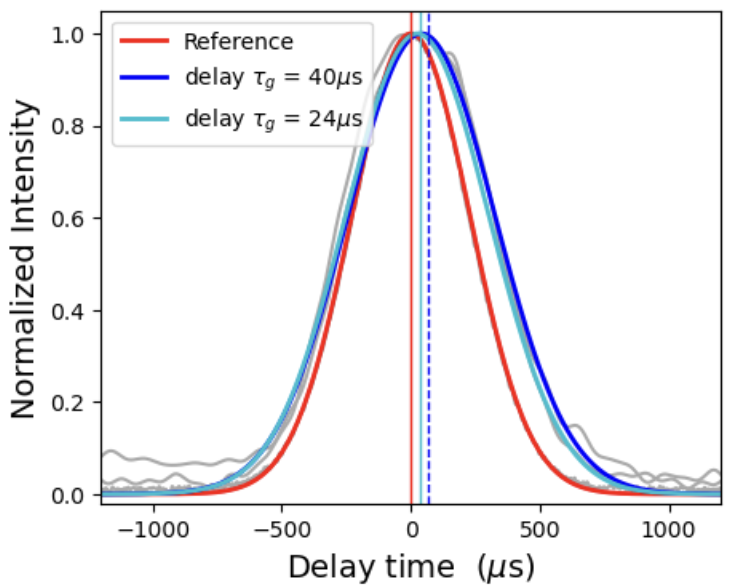}  
\caption{\label{fig:fig3} Probe and reference pulses shown on the oscilloscope. The red and blue curves are Gaussian fits of the signals for two different extinction ratios between the control and probe beams, achieved by adjusting the probe beam power.}
\end{figure}

To characterize the slow-light effect, we measured the group delay (that is, the time delay) \(\tau\) by locking the probe-beam frequency to the EIT transmission peak. The probe beam was modulated into short pulses with the EOM, and the transmitted signal was detected with a photodiode connected to an oscilloscope. The relative delay between the reference and slowed pulses was analyzed by first fitting the transmitted pulses to a Gaussian curve \(\tau\). We measure the delay \(\tau\) between the probe pulse traveling through the Rb cell and the reference pulse traveling through air. Group delay measurements of 24 \(\mu\)s and 40 \(\mu\)s are shown in Fig.~\ref{fig:fig3}, corresponding to group velocities of \( v_g \approx 3125 \, \text{m/s} \) and \( v_g \approx 1875 \, \text{m/s} \), respectively. 

These values confirm the steep dispersion induced by EIT, demonstrating the effectiveness of non-degenerate Zeeman EIT in achieving significant group velocity reduction. The measured delays provide a crucial parameter for investigating the transverse light-drag effect.

\section{Theoretical Framework and Experimental Setup}

To explore the invariance of the transverse drag effect, we consider two key scenarios: (1) the Rb cell moves transversely with respect to a stationary probe beam, and (2) the probe beam moves transversely with respect to a stationary Rb cell. In our context, we distinguish reference frames from experimental scenarios: the lab frame refers to the coordinates fixed to the optical table and measurement apparatus, while the scenarios describe two physically different configurations implemented in the lab. In the first case, the displacement $\Delta x'$ in the probe beam frame (primed coordinates), where the Rb cell moves with transverse velocity $v$, is given by $\Delta x' = v \tau$, where $\tau$ is the group delay induced by the interaction of light and the medium.  Transforming to the lab frame (unprimed coordinates), where the probe beam is stationary, we use the Galilean transformations:

\begin{equation}
x = x' + vt', \quad t = t'.
\label{eq:galilean_transform1}
\end{equation}

Thus, the displacement remains $\Delta x = \Delta x' = v \tau$.

In the second scenario, where the probe beam moves with velocity $v$ relative to the stationary Rb cell, the displacement in the lab frame is $\Delta x = v \tau$. To transform this into the probe beam’s frame, we use:

\begin{equation}
x' = x - vt, \quad t' = t.
\label{eq:galilean_transform2}
\end{equation}

Again, the displacement observed in the probe frame is $\Delta x' = \Delta x = v \tau$. While this shows that, in the non-relativistic regime, the displacement $\Delta x$ is invariant between frames, direct measurement in both configurations is necessary to confirm this under real experimental conditions. Potential experimental imperfections, such as optical misalignments, residual Doppler effects, and variations in slow-light conditions, require a direct measurement of transverse drag in both moving frames. The following sections describe the methodology used to test this invariance in a controlled setup.

\begin{figure*}[t]
\centering
\includegraphics[width=1.0\linewidth]{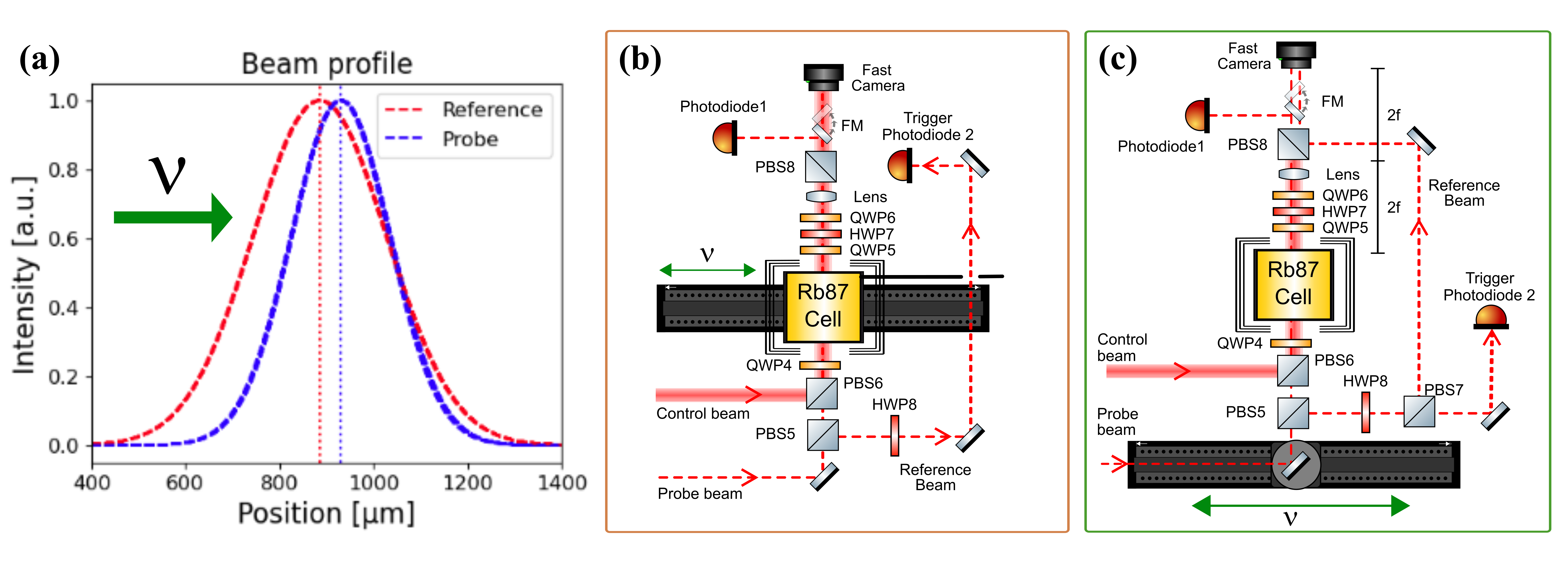}%
\caption{\label{fig:fig4}: Left: (a) Transverse drag enhanced by slow light in a medium moving at $v=350 \text{ }\mathrm{mm/s}$; the green arrow indicates the direction of motion. Right: Experimental configurations to investigate transverse light-drag in moving media. Both depicted setups display only the part of the setup used to measure transverse light drag using slow-light induced by non-degenerate Zeeman EIT. (b) Setup for a moving medium with a stationary probe beam. The trigger mechanism consists of an extension barrier attached to the moving medium, and an iris placed such that the reference beam is temporally aligned with the probe beam exactly when the probe passes through the center of the Rb cell. (c) Setup for a moving probe beam with a stationary medium; we used a PBS to extract part of the moving probe beam to trigger the camera, exactly when the probe passes through the center of the Rb cell.}
\end{figure*}

The experimental setup, depicted in Fig.~\ref{fig:fig4}, implements these two configurations: (1) a moving medium with a stationary probe beam, and (2) a moving probe beam with a stationary medium.

Before measuring the light drag, the time delay produced by the EIT slow-light effect was recorded, as detailed in the previous section. This step is crucial for maintaining an optimal extinction ratio between the control and probe beam powers, ensuring effective slow-light operation. After measuring the group delay using pulsed probe light generated by the electro-optic modulator (EOM), the EOM was turned off to switch to continuous-wave operation for the photon-drag measurements. Since turning off the EOM slightly altered the probe beam power, it was then readjusted to maintain the desired extinction ratio between the probe and control beams.

\subsection{Moving Medium Frame}

The setup to measure transverse drag with a moving medium and a static probe beam is shown in Fig.~\ref{fig:fig4}b. Here the Rb cell is mounted on a translation stage (Thorlabs DDS600) allowing the cell to move sideways in a controllable fashion. Before the control and probe beams are recombined, we use PBS5 to split off a portion of the probe beam that will be used as a trigger mechanism that will ensure accurate timing of the measurement. We attached an iris to the Rb cell, which is fixed and synchronized to let this portion of the probe beam pass through exactly when the probe beam is at the center of the Rb cell.

Data collection involves recording 100 frames, with 50 frames for the probe moving to the right and 50 for the probe moving to the left. This process is repeated for each speed, ranging from $v = -350$ mm/s to $v = 350$ mm/s. 
Measurements are conducted under both EIT conditions, when the control and probe satisfy the two-photon resonance (that is, on-resonance), and under off-resonant or non-EIT conditions, where we also blocked the control beam to prevent leakage.

Data analysis involves averaging the centroid positions of the probe and reference beams and subtracting one from the other to obtain the effective distance, $\Delta x_0$, between the beams. Additionally, the offset distance, $\Delta x'$, is measured for the off-resonance case. To determine the true displacement, $\Delta x$, we subtract $\Delta x'$ from $\Delta x_0$. 

\subsection{Moving Probe Frame}

In the case of a moving probe beam and stationary medium, the mirror that directs the probe beam to the Rb cell is mounted on the translation stage, allowing the reflected beam to move sideways as shown in Fig.~\ref{fig:fig4}c. This time we use the portion of the moving probe beam created by PBS5 for two purposes: (1) to create a beam that travels outside the Rb cell and that serves as a reference, and (2) to trigger the fast camera, which captures both the transmitted probe and reference beams, displaying them separately on the camera.

We follow the same procedure as described above, to obtain three distinct measurement sets corresponding to delays of $24$, $30$, and $40 \mu s$, as shown in Fig.~\ref{fig:fig5}b. The data analysis follows a similar approach, where we average the centroid position of the beam and determine the displacement by subtracting the off-resonance case from the on-resonance case.

\section{Results and Discussion}

Our experiment successfully demonstrates Galilean invariance of transverse light drag using non-degenerate Zeeman electromagnetically induced transparency (EIT) in rubidium vapor. By comparing scenarios where either the probe beam or the medium (Rb cell) moves transversely, we found that the displacement $\Delta x$ is the same in both situations, as predicted by the principle of Galilean invariance. This linear relationship with velocity confirms the theoretical predictions, validating the experimental approach and establishing a robust framework for investigating light-matter interactions in moving media.

\begin{figure*}[ht]
\centering
\includegraphics[width=1.0\linewidth]{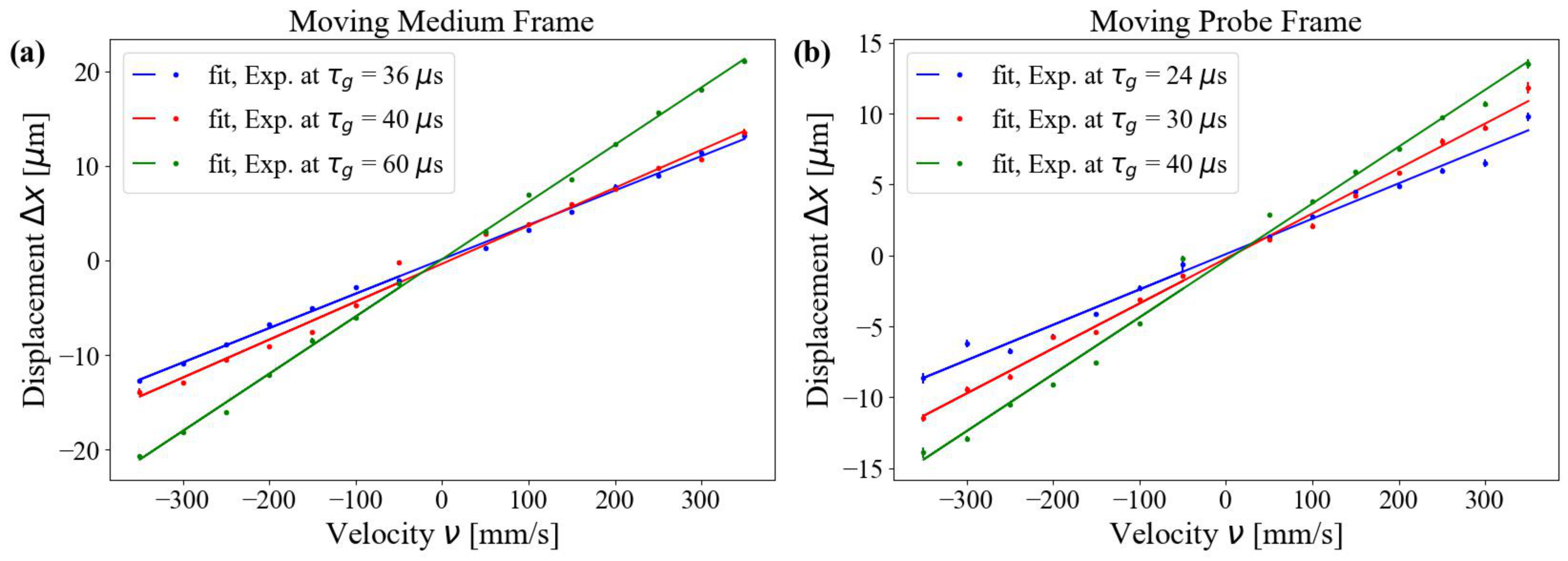}%
\caption{\label{fig:fig5}Transverse drag measurement versus velocity, for the cases when the motion is in the left direction (negative speeds) and in the right direction (positive speeds). (a) Moving medium frame: The linear fits are in good agreement with expected values measured of time delays of $36$, $40$ and $60 \mu s$, within less than 10\%. (b) Moving probe frame: The linear fits are also in good agreement with that expected from the measured time delays of $24$, $30$ and $40 \mu s$, within less than 10\%. The data points are the means of 50 measurements, and the small error bars are their standard deviations. }
\end{figure*}

Figure~\ref{fig:fig5} presents the measured displacement versus velocity for both configurations: the moving medium (Fig.~\ref{fig:fig5}a) and the moving probe beam (Fig.~\ref{fig:fig5}b). The linear fits closely match the measured values of the group delay, with discrepancies below 10\%. This agreement supports the expected relationship $\Delta x = v \tau$, confirming the equivalence of the transverse drag effect in both frames and thereby demonstrating Galilean invariance.

To evaluate the sensitivity of our setup, we determined the minimum measurable velocity using:
\begin{equation}
V_{\text{sensitivity}} = v_g \frac{\Delta x_{\text{err}}}{L},
\end{equation}
where $\Delta x_{\text{err}} = 0.15 \, \mu\text{m}$ is the standard deviation in displacement measurements, $L = 7.5 \, \text{cm}$ is the Rb cell length, and $v_g = 500 \, \text{m/s}$ corresponds to the slowest group velocity achieved in our experiment, with a group delay of $150 \mu s$. The resulting velocity sensitivity is $0.001 \, \text{m/s}$, demonstrating the capability of our system to resolve small velocity variations.

The maximum measurable transverse velocity is limited by the requirement that the beam remains within the aperture of the Rb cell between consecutive camera frames. Given a beam diameter \( d \) and a cell aperture \( D \), the maximum allowable displacement per frame is \( D - d \). Multiplying this by the camera frame rate \( f \) yields:
\begin{equation}
V_{\text{max}} = (D - d) f,
\end{equation}
with \( D = 2 \, \text{cm} \), \( d = 1 \, \text{mm} \), and \( f = 55 \, \text{Hz} \), resulting in \( V_{\text{max}} = 1045 \, \text{mm/s} \). This defines the upper bound of the system’s dynamic range.

The effectiveness of EIT-based light-drag measurements is strongly influenced by the optical depth (OD) of the medium, which determines the background absorption level, the contrast of the transparency window, and the steepness of the dispersion curve. These factors directly impact the group velocity and the sensitivity to transverse beam shifts caused by motion. We estimate the OD through knowledge of the oscillator strength of the Rb transition, the atomic number density, and the decoherence rate of the transition $\gamma_{13}$, which is influenced by both Doppler broadening ($\Gamma_D$) and the collisional dephasing rate ($\gamma_{\text{coll}}$) due to the buffer gas. The contrast $C$ of the EIT transparency window, which we can use to quantify the theoretical transparency, is given by:
\begin{equation}
C = \frac{|\Omega_c|^2}{\gamma_{13} \gamma_{12} + |\Omega_c|^2},
\end{equation}
where $\Omega_c$ is the Rabi frequency of the control field, $\gamma_{13}$ is the optical decoherence rate, and $\gamma_{12}$ accounts for decoherence between the ground states. This expression shows how higher control field intensity improves the EIT contrast, as also discussed in ~\cite{kenneth_derose_producing_2020,ran_finkelstein_practical_2023}

The total optical decoherence rate is determined primarily by Doppler broadening ($\Gamma_D$) and collisional dephasing ($\gamma_{\text{coll}}$). For our 30 Torr He buffer gas, the collisional broadening is estimated as:
\begin{equation}
\gamma_{\text{coll}}/2\pi = 30 \times 2.5 \, \text{MHz} \approx 75 \, \text{MHz},
\end{equation}
while the Doppler broadening contributes approximately 250 MHz, resulting in a total decoherence rate of:
\begin{equation}
\gamma_{13}/2\pi = \gamma_{\text{coll}} + \Gamma_D \approx 325 \, \text{MHz}.
\end{equation}

With these values, the optical depth is calculated as:
\begin{equation}
\text{OD} = \alpha_0 L,
\end{equation}
where  $\alpha_0 = (3 \lambda^2 \Gamma N) (8 \pi \gamma_{13})^{-1}$ is the on-resonance absorption coefficient.

Using experimental parameters \( \lambda = 794.98 \, \text{nm} \), a control beam intensity of 0.8–2.1 mW/cm\(^2\), and an atomic number density of \( N = 3.8 \times 10^{13} \, \text{m}^{-3} \) at \( 65^\circ\text{C} \), we estimate an OD of approximately 68. Based on the associated Rabi frequency and the contrast expression above, this corresponds to a theoretical transparency (i.e., EIT contrast) in the range of 71–86\%. While this expression provides insight into the dependence of EIT contrast on various parameters, experimental values often deviate due to additional broadening mechanisms and imperfections in the system as discussed by DeRose et al \cite{kenneth_derose_producing_2020}. Our observed transparency ranged between 20–25\%, likely due to environmental fluctuations and optical misalignments affecting the slow-light conditions.

Despite these decoherence effects, the full-width at half-maximum (FWHM) of the EIT window remained between 7.65 kHz and 10.33 kHz, indicating stable slow-light conditions and reinforcing the robustness of our setup for precision velocity measurements. Investigating light propagation in arbitrary velocity fields remains a fundamental challenge in physics. The slow-light enhancement in our setup enables precise laboratory studies of these effects, with applications in inertial sensing, fluid dynamics, and optical image processing.

The drag effect and velocity resolution could be significantly enhanced through use of optical storage, where the probe pulse is mapped into atomic states and later retrieved after the transverse motion of the medium. Recent work by Ahmadi et al. (2024) suggests that storing light in EIT media can extend sensitivity beyond typical slow-light delays, with storage times reaching milliseconds, potentially amplifying the effect by 1 to 2 orders of magnitude. These advancements could enable high-sensitivity velocimetry for quantum metrology, remote sensing, and tracking of moving objects in turbulent environments.

\section{Conclusion}

Our results provide experimental verification of the Galilean invariance of transverse light drag, reinforcing theoretical predictions. This study establishes a robust optical platform for investigating light-drag effects in moving media, with potential applications in quantum optics, nonlinear photonics, and precision velocimetry. A key advantage of this approach is its capability of performing stand-off measurements, distinguishing it from traditional velocimetry techniques. This feature is particularly relevant for velocity sensing in hazardous environments where direct sensor placement is impractical. Additionally, this approach may provide a valuable tool for probing light-matter interactions under gravitational fields, as explored by Dressel et al.~\cite{justin_dressel_gravitational_2009} and other groups ~\cite{paul_dixon_ultrasensitive_2009, jordan_gravitational_2019} where they used interferometric weak value techniques to amplify small deflections. Furthermore, this technique could be extended to resolve local velocity gradients in spatially varying or turbulent flow fields, as recently demonstrated via holographic reconstruction of slow-light wavefronts in moving cold-atom clouds~\cite{yuzhuo_wang_imaging_2021}.

Our results highlight the essential role of experimental optimization in translating theoretical principles into real-world measurements. Rather than a direct bridge from theory to experiment, each setup requires careful integration of multiple factors to achieve optimal performance under specific conditions. By developing two distinct configurations to investigate transverse drag in moving media, we provide a framework for future studies to refine optical depth control and pave the road for system miniaturization for enhanced practicality in the context of light-drag measurements.

\section{Acknowledgements}

We gratefully acknowledge financial support by the US Office of Naval Research award N00014-19-1-2247 and MURI award N00014-20-1-2558, and US National Science Foundation Award 2138174. The portion of the work performed at the University of Ottawa was supported by the Canada Research Chairs program under award 950-231657, the Natural Sciences and Engineering Research Council of Canada under Discovery Grant RGPIN/2017-06880, and the Canada First Research Excellence Fund Award 072623.

We deeply appreciate invaluable discussions with Prof. Nir Davidson, and Prof. Ofer Firstenberg. We thank Dr. A. Nicholas Black, Dr. Giulia Marcucci, and Dr. Jerry Kuper for their help during the initial setup.

E.S.A.R. and L.D.N. contributed equally to this work.

\bibliography{references01}

\end{document}